# Measuring Chiral Phonons

**Rahul Rao,[1,*] Hanyu Zhu,[2,*] Nicholas A. Kotov[3,*] Thuc T. Mai,[1,4] Maria F. Muñoz,[5] Dali Sun,[6] Jun Liu,[7] Wonjin Choi,[8] Renee R. Frontiera,[9] Angela R. Hight Walker[4]**

[1]Materials and Manufacturing Directorate, Air Force Research Laboratory, Wright-Patterson AFB, Ohio 45433, United States

[2]Department of Materials Science and NanoEngineering, Rice University, Houston, Texas 77005, United States

[3]NSF Center for Complex Particle Systems (COMPASS), Department of Chemical Engineering, Department of Materials Science and Engineering, Department of Biomedical Engineering, Biointerfaces Institute, University of Michigan, Ann Arbor, Michigan 48109, United States

[4]AV Inc., Arlington, VA 22202

[5]Quantum Measurement Division, Physical Measurement Laboratory, National Institute of Standards and Technology, Gaithersburg, Maryland 20899, United States

[6]Department of Physics, North Carolina State University, Raleigh, NC 27695, United States

[7]Department of Mechanical and Aerospace Engineering, North Carolina State University, Raleigh, NC 27695, United States

[8]Lawrence Livermore National Laboratory, Livermore, California 94550, United States

[9]Department of Chemistry, University of Minnesota, Minneapolis, MN 55455, United States

[*]e-mail: rahul.rao.2@us.af.mil, Hanyu.zhu@rice.edu, kotov@umich.edu

**ABSTRACT**

Chiral phonons are quantized vibrations where the atomic motion in a solid breaks improper rotation symmetries. In many cases, chiral phonons possess angular momenta and are therefore selective to circularly polarized light. Both fundamental and applied research efforts on chiral phonons have been gaining increasing attention owing to their importance in a variety of fields including spintronics, spin-selective chemical reactions, thermal transport, quantum information processing and biosensing, where the bi-directional spin-lattice coupling enabled by chiral phonons can be harnessed in new ways, and potentially lead to new functionalities. Thus far, the studies of chiral phonons across diverse materials platforms have evolved largely independently within these fields, but the experimental techniques are often interrelated. In this perspective, we present a detailed description, as well as advantages and disadvantages of the current approaches for experimentally measuring chiral phonons in chiral and achiral materials. We conclude with a discussion of new methods for measuring chiral phonons. Ultimately, this work seeks to offer an experimental guide for systematically investigating the properties of chiral phonons in various materials systems and applications.

**Key points:**

- Chiral phonons are quantized lattice vibrations in which the atomic motion breaks improper rotational symmetry, often associated with rotation around their equilibrium positions and intrinsic angular momentum, thus making them sensitive to circularly polarized light.
- Fundamental knowledge about chiral phonons and their experimental registration is essential for diverse scientific fields, including spintronics, spin-selective chemistry, thermal transport, quantum information processing and biomedical research.
- Chiral phonons are naturally present in crystals of biomolecules, such as left- and right-handed amino acids, as well as crystals of achiral molecules with lattices belonging to chiral space groups; chiral phonons can also emerge transiently in achiral materials through symmetry breaking.
- A variety of techniques including optical rotatory dispersion, circular dichroism, vibrational optical activity (including vibrational circular dichroism and Raman optical activity), resonant inelastic X-ray scattering, and time-resolved spectroscopy, provide complementary means of detecting chiral phonons in steady-state and transient regimes.
- While for some techniques such as terahertz circular dichroism, chiral phonons can result in giant optical activity, the magnitude of spectroscopic signatures corresponding to chiral phonons may be orders of magnitude weaker for other techniques.
- The authors propose an experimental guide toward more sensitive detection methods and cross-disciplinary collaboration to exploit these quasiparticles in future technologies.

**Website summary**: This perspective explains the presence of chiral phonons carrying angular momentum in chiral and achiral materials, summarizes experimental methods for detection, including optical circular dichroism, incoherent scattering, and coherent excitation techniques, and highlights emerging interdisciplinary applications and challenges.

**Introduction**

Phonons are quantized vibrations in solids and correspond to collective motions of atoms in crystals. Traditionally, they have been considered to possess only linear momentum but not angular momentum (AM) like electrons do. However, experimental and theoretical observations of spin-phonon coupling in bulk three-dimensional (3D) materials led to the discovery of non-zero AM in phonons[1–5]. This concept was then also extended to non-magnetic two-dimensional (2D) materials lacking inversion symmetry, where chiral phonons were predicted and experimentally observed at high symmetry points of their Brillouin zones[6–8]. Recently, similar effects were also observed in one-dimensional (1D) crystals[9,10].

Broadly speaking, **chiral phonons are quantized vibrations where the motion of an atom or a group of atoms breaks all improper rotation symmetries in a solid, such as rotation perpendicular to the direction of vibrational wave propagation**. Chiral phonons may be present in both chiral and achiral materials, as discussed in Box 1. In non-chiral materials, two kinds of chiral phonons are distinguishable[11]. The first are geometric chiral phonons, which do not exhibit AM and where the atomic displacements locally break mirror symmetry, inducing a transient chiral distortion. These phonons may be observable through circular dichroism. The second are axial chiral phonons, which possess AM, hence their name follows the description of an AM vector[11,12]. The rotational motion of axial chiral phonons imparts an intrinsic net AM and intrinsic handedness to such oscillations. In turn, this leads to a variety of novel collective phenomena such as the phonon Hall,[13,14] Einstein-de Haas,[15,16] Zeeman[17,18] and spin Seebeck effects[19], nanoscale band narrowing[20], and resonance coupling between chiral phonons and plasmons[21]. For a broader overview of the fundamental theory and phenomenology of chiral phonons, we direct the reader to existing general reviews[8,22,23].

Despite rapid theoretical progress, a comprehensive synthesis of the experimental toolkit required to probe these modes is lacking. Recent advances in spectroscopic methods now permit direct spatiotemporal and momentum-resolved access to chiral phonons. A focused review of these measurement techniques is therefore particularly timely to help standardize methodologies, navigate experimental selection rules, and establish practical guidelines for measuring chiral phonons. In this Technical Review, we evaluate current experimental methods used to identify and characterize geometric and axial chiral phonons across chiral and achiral platforms. We structure this review by first examining steady-state probe techniques, then progress to transient time-resolved spectroscopies. Some current application avenues are also discussed in Box 2. Finally, we conclude with our thoughts on future prospects for measurement tool development.

**Box 1. Materials that host chiral phonons**

Starting with Louis Pasteur's 1848 discovery that separated enantiomers of ammonium tartaric acid rotated plane-polarized light in different directions[24], the number of chiral materials has exploded to encompass a wide range of length scales and hierarchies, from sub-nanometer atomic clusters to a large variety of nanostructures to supramolecular assemblies[25–28]. Since chiral phonons propagate in crystal lattices, it is important to point out that the chirality of a material is intimately linked to its structure through the crystallographic space group. Of the 230 possible space groups, 65 of them are chiral, lacking an inversion center or a mirror plane, which implies that there are several thousand potential chiral materials[29,30]. Although the remaining 185 space groups are achiral, it is possible for the atomic displacements of a phonon mode to locally break the achiral symmetry, inducing a transient chiral distortion[12]. This symmetry-breaking can be initiated through interactions with photons, magnons, or comparable excitations. Furthermore, non-zero linear and/or AM of phonons can break mirror symmetry, allowing chiral phonons in any solids, including non-crystalline materials. For example, non-propagating chiral phonons at high symmetry points in achiral 2D materials were first reported theoretically in 2015,[6] then experimentally observed in $WS_2$ in 2018[7]. In these materials the phonon mode involves the rotational motion of the atoms within the 2D honeycomb lattice. Also in 2018, chiral phonons were experimentally observed in nanoparticles of $Co_3O_4$ taking advantage of their Raman optical activity (ROA)[5]. Note that the origin of the chiral phonons in this case are chiral distortions produced by organic ligands on an otherwise achiral crystal lattice. Also, in ionic achiral lattices such as PbTe, the rotational motion of the ions can induce magnetic moments on the order of 10 $\mu_B$, which is measurable using ultrafast spectroscopic methods[18,31,32].

All crystals belonging to the 65 chiral space groups can support the propagation of chiral phonons. However, whether the lattices in these space groups reveal many of the effects related to the coupling of chiral phonons to other quantum phenomena, spin, and thermal transport, is a more difficult question to answer. A recent study attempted to answer this question by a high-throughput symmetry-based analysis of phonon AM across all 230 space groups[303233]. Of the 11,614 crystalline compounds screened, they identified chiral phonon modes in 2738 materials. The results are compiled into an open-access website (Chiral Phonon Materials Database, https://materialsfingerprint.com). To highlight the structural complexity in materials that host chiral phonons, in **Fig. 1** we show left- and right-handed counterparts of some exemplar chiral crystals. From relatively simple chiral crystals like Te and $\alpha$-quartz, structural chirality can extend to hybrid organic-inorganic perovskites and more complex biomolecules. The vast number of materials with chiral phonons heralds an exciting time ahead for experimentalists to characterize them and to potentially manipulate them through their interactions with other quasiparticles in the systems. Moreover, the list of chiral materials is ever expanding[34,35] with innovations in crystal growth and nanostructure assembly, and it is possible for many of these chiral systems to host propagating chiral phonons.

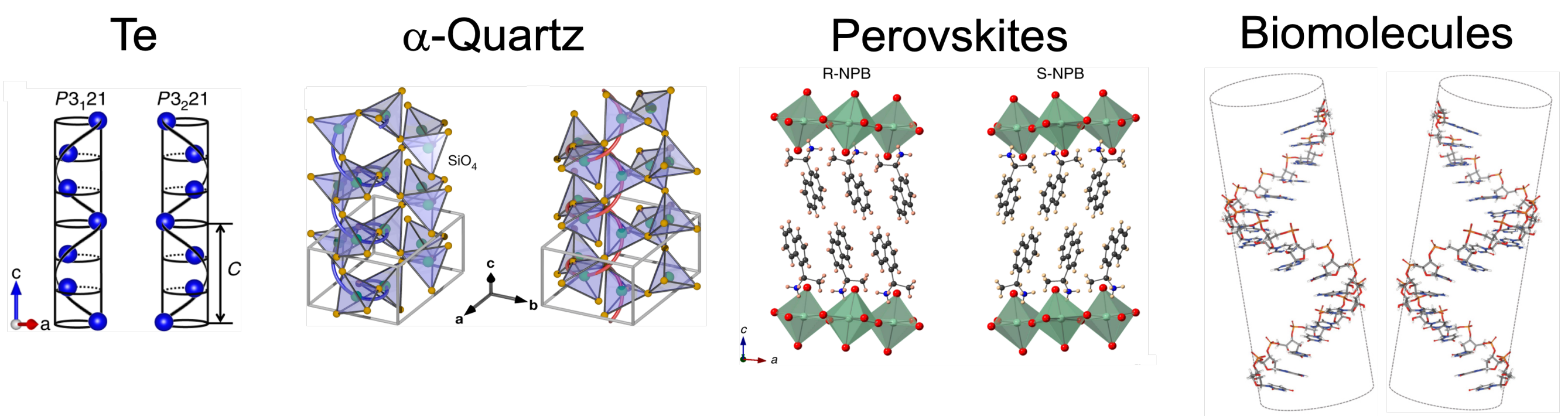

**Figure 1.** Examples of left- and right-handed chiral materials hosting chiral phonons. The crystal structures going from left to right showcase an increasing degree of structural complexity.

**Measurement techniques**

In the following sections we discuss current approaches to experimentally measure chiral phonons in chiral and achiral materials. We have broadly divided the sections into steady state and transient techniques, which are discussed in terms of the instrumentation as well as the breadth of information provided on the chirality of phonons in both chiral and achiral materials.

*Steady state techniques*

Considering that chiral materials are very sensitive to interactions with circularly polarized light, optical characterization methods are the most popular for measuring and quantifying chirality. The two techniques that historically have been the benchmarks are optical rotatory dispersion (ORD) and circular dichroism (CD). The latter colloquially often refers to CD spectroscopy, measured for visible range wavelengths and is also known as electronic circular dichroism (ECD). ORD measures the change in the optical rotation of linearly polarized light passing through a material as a function of wavelength while CD measures the differential absorption of right and left circularly polarized light (RCP and LCP, respectively) by a chiral material. In the case of CD related to transitions involving excited electronic states, it is important for the material to contain a light absorbing chemical structure, *i.e*. a molecule, cluster, or nanoparticle, thus limiting the technique to the UV-visible wavelength range. On the other hand, ORD can be used at wavelengths where the material does not exhibit significant absorption. Thus, each method has its own advantages and disadvantages, with CD being the more popular technique and affording high resolution understanding of molecule conformation, while ORD is an older technique that is used heavily in the pharmaceutical industry for determining absolute configurations of smaller, well-defined chiral molecules[36–38]. Note that physically and mathematically the spectra obtained with CD and ORD are not independent: the complex rotatory strength has real (dispersion) and imaginary (absorption) components, and the Kramers–Kronig relations link them. In practice this means the ORD curve can be calculated by integrating the CD curve over wavelength, and vice versa.[39] Because CD has narrow absorption bands, it generally offers higher spectral resolution than ORD, whose dispersive features are broader; however, ORD can be measured at wavelengths where the sample does not absorb, making it useful for regions outside the absorption band.

The foundations of the ORD and CD spectroscopies operating in the visible range of electromagnetic waves can be extended towards vibrational transitions in chiral materials. Vibrational optical activity (VOA) is a spectroscopic measure of the differential absorption of a chiral molecule for RCP and LCP excitation during a vibrational transition. The two standard methods for measuring VOA are infrared (IR) vibrational circular dichroism (VCD) and Raman optical activity (ROA). While standard IR or Raman scattering spectra from chiral enantiomers are identical, their VCD or ROA spectra are opposite in sign (mirror images of each other) owing to the differential response to LCP and RCP excitation. The VCD and ROA spectra therefore often look often very similar to CD spectra. VCD and ROA measurements were first reported in the early 1970s[40,41], and have since then progressed significantly in terms of improvements in sensitivity and instrumentation[42,43]. These techniques also benefit from strong theoretical support where density functional theory (DFT) calculations of VCD and ROA spectra from chiral molecules have enabled the determination of absolute configurations[44]. The primary difficulty in measuring VOA is that its signals are four to five orders of magnitude smaller compared to IR absorption and Raman scattering[41,45]. In order to measure these small signals, the incident laser is modulated between the LCP and RCP states. In ROA the scattered light is collected in separate channels on a detector while the VCD is obtained by amplification and phase-sensitive detection of the modulated signal[46,47]. These measurements may take several minutes to hours but yield valuable molecular insights into the conformations of chiral molecules. It should be noted that the order of magnitude of the Raman scattering intensities and their differences can vary between different molecules, and can be significant in single crystals[48].

Raman spectroscopy

Raman scattering is the inelastic scattering of light by matter. The process involves the absorption of photons by a material, which excites its molecules to a virtual state. The molecules then relax down to a higher energy vibrational state, emitting photons with lower energies (Stokes process). For molecules already in higher vibrational states, the scattered light gains energy (anti-Stokes process). The Raman scattered light provides information about the vibrational modes, electronic states, and magnetic excitations. As mentioned above, the instrumentation for ROA is often specialized, but recent measurements of chiral phonons have used conventional Raman spectrometers with circularly polarized excitation. In the rest of this section, we will highlight the main principles behind a circularly polarized Raman scattering experiment. The commonality between the circularly polarized Raman measurements employed in recent experiments and the highly specialized ROA instruments, as well as a few key differences, is mentioned. Finally, we will provide a brief discussion of the current understanding of Raman scattering of chiral phonons.

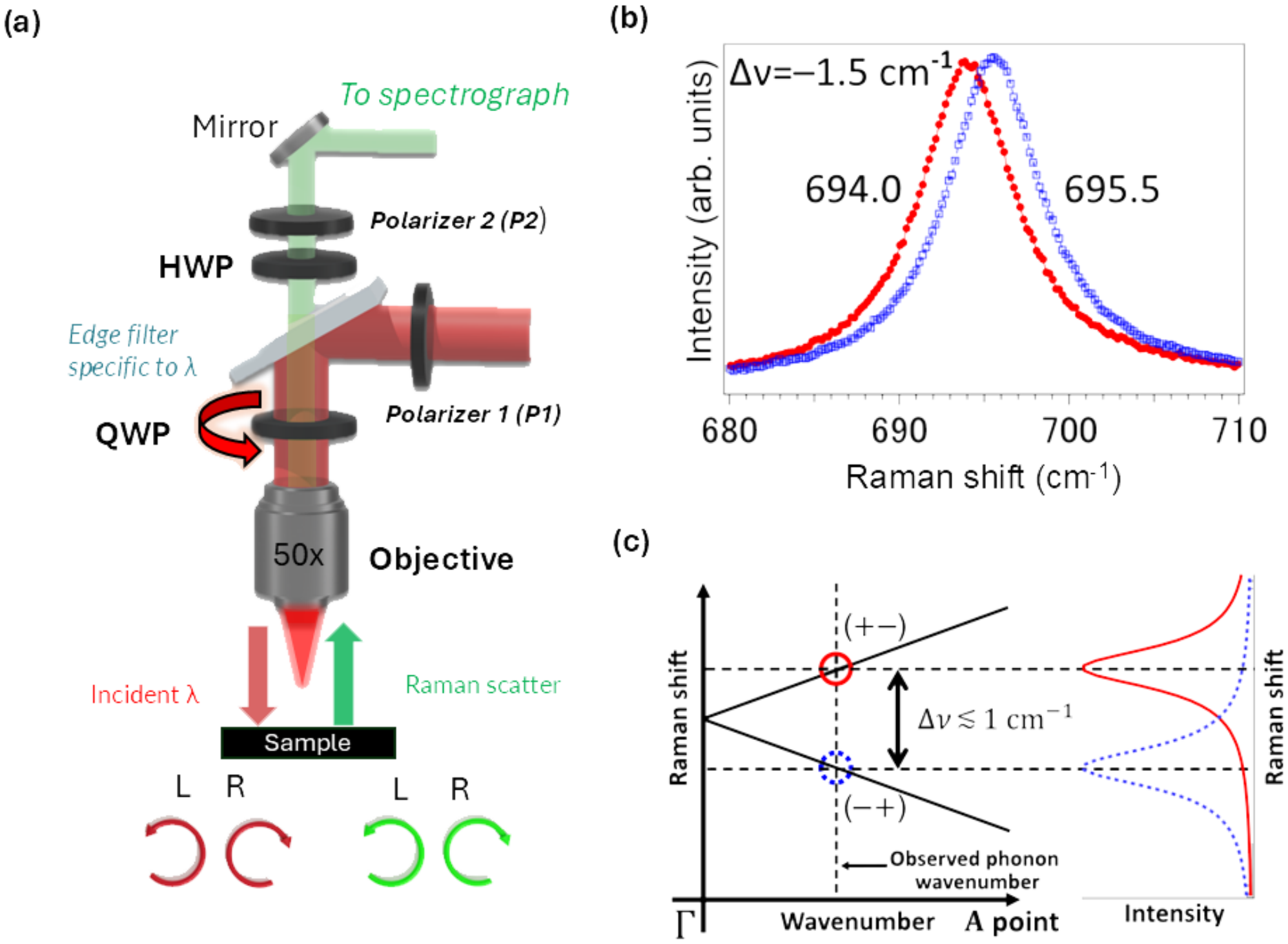

**Figure 2. Circularly polarized Raman spectroscopy of chiral phonons**. (a) Typical commercial design of a Raman spectrometer coupled to a microscope. (b) Schematic of the dispersion of chiral phonon near the Brillouin zone center (Γ). Differences in AM result in a splitting of peak frequencies at a small but finite wavenumber away from the Γ point, as depicted by the peaks on the right. (c) Cross-circularly polarized Raman Spectra (LR and RL configurations) from left-handed α-quartz. (b) and (c) taken from Ref. [49].

A typical Raman spectrograph makes use of a diffraction grating and a scientific-grade CCD detector. It is also crucial to consider the effect of the excitation energy on the CD and optical activity in the material being studied. Resonance with an electronic transition will significantly affect the measured ROA signal, making it difficult to distinguish the vibrational contribution. Other than these considerations, the most important components for the circularly polarized Raman measurement involve the polarization states of the excitation laser and of the scattered light. Since there are two possible circular polarization states, left (L) and right (R), there are four combinations considering the incident and scattered polarizations. These are LL, RR, LR, and RL, where the first (second) letter corresponds to the incident (scattered) polarization. These polarization configurations can be achieved by the use of polarizers (P1 and P2) and waveplates (QWP and HWP) as shown in **Fig. 2a**[50]. It is important to note that the Raman intensities measured from these polarization configurations, $I_{LL}$, $I_{RR}$, $I_{LR}$, $I_{RL}$, provide the complete polarization response of a sample in a given scattering plane. From these measurements, the selective absorption of RCP or LCP light by chiral phonon modes can be measured with the ROA ($I_{RL} - I_{LR}$) and the circular intensity difference, CID = $(I_{Right} - I_{Left})/ (I_{Right} + I_{Left})$. Here $I_{Right}$ and $I_{Left}$ represent the measured Raman intensity under different circular polarization conditions. The most general approach is to add all components for which the incident light is right- or left-circularly polarized, i.e., $I_{Right} = I_{RL} + I_{RR}$, and $I_{Left} = I_{LR} + I_{LL}$. However, these definitions can be simplified to specific cases, considering only the co-circular or cross-circular components, for example, CID= $[(I_{RL} - I_{LR})/ (I_{RL} + I_{LR})]$[45,51]. We note that the ROA as defined here can be considered as a simplified version of the traditional ROA measurement, which, as described above, is a dynamic measurement involving specialized instrumentation.

Other than intensities, a distinct feature and the main focus of chiral phonon studies in chiral materials is the pseudoangular momentum (PAM)-induced frequency splitting of a phonon mode[10,49,52–54]. In short, to conserve AM during the Raman scattering process[55], the spin AM of the incident circularly polarized photon is transferred to the lattice, exciting a chiral phonon. Depending on the chosen circular polarization configuration (LR or RL), the transferred AM can be used to selectively excite chiral phonon modes with opposite PAM, at small but finite wavevectors away from the Brillouin zone center (**Fig. 2b**). This manifests as the splitting of phonon mode frequencies, which can be experimentally measured with different circular polarization configurations, e.g. $I_{LR}$ vs. $I_{RL}$[49,53,54]. An example of this is plotted in **Fig. 2c**, which shows the 1.5 cm$^{-1}$ splitting of a degenerate phonon mode in $\alpha$-quartz, measured using cross-circularly polarized Raman spectroscopy[49]. We note that phonon frequency splitting in achiral materials has also been observed upon the application of an external magnetic field[56–59]. In these materials, the phonons gain AM by coupling with a spin-flipping transition, such as crystal-field levels, spin-orbital splitting, magnon excitations, etc[60–65]. Such coupling between spin and electronic states has also been observed in neutron scattering and thermal transport measurements[13,66–71].

Experimentally, careful considerations must be made in a circular polarization-sensitive measurement. Characterizing the excitation beam to check for any ellipticity in the circular beam ensures a higher probability of success. Because the Raman scattered light can cover a large frequency range, the polarization response of all the optical components in such a system must be uniform in a sufficiently broad bandwidth[50]. Furthermore, to correctly measure the frequency splitting, circularly polarized Raman spectra should be collected by exciting the chiral material along its crystallographic chiral axis (e.g. the *c* axis of Te). Off-axis excitation at an angle may result in complications such as elliptic polarization and linear birefringence[72], which may obscure any frequency splitting in peaks and/or variations in Raman peak intensities (ROA). Lastly, while a splitting of peak frequencies between 1 – 2 cm$^{-1}$ is measurable using commercial Raman spectrometers, circularly polarized Raman studies on chiral and achiral materials have revealed much smaller frequency splits around 0.1 - 0.2 cm$^{-1}$, [48,73,74] which are much lower than the spectral resolution of many instruments. Interestingly, the modes that exhibited very small frequency splitting showed strong signatures of ROA[48,74,75] and DFT calculations showed that some modes did indeed possess PAM[48,74]. A recent circularly polarized Raman study on enantiomers of CoSi also reports that doubly degenerate *E* symmetry modes exhibit ROA while triply degenerate *T* symmetry modes exhibit splitting of peak frequencies[76]. Here, chirality lifts the degeneracy to produce frequency splitting, whereas axial symmetry breaking produces the ROA. Thus, relying on either differences in peak intensities or in peak frequencies are by themselves not sufficient to provide a definitive fingerprint of a chiral phonon. DFT calculations of PAM are critical in distinguishing between axial and geometric chiral phonons.

A related inelastic scattering technique, resonant inelastic X-ray scattering (RIXS), with its much higher momentum transfer can resolve larger energy splitting away from the $\Gamma$ point[77], helping mitigate the resolution limitations of Raman spectrometers. RIXS exploits X-ray photons resonant with core electrons with a keV energy scale[78], and has been employed successfully to measure chiral phonons in $SiO_2$ and $LiNbO_3$[77,79]. The main limitation of the technique is that the scattering in RIXS is extremely weak, necessitating high incident photon fluxes to obtain enough measurable scattered photons[80]. Other limitations include availability of instruments with sufficient energy resolution as well as a lack of control over the polarization state of the scattered light[81].

## Chiroptical Terahertz Spectroscopy

Terahertz (THz) circular dichroism (TCD) and THz optical rotational dispersion (TORD) both extend the concepts described above for CD, VCD and ROA into the terahertz (THz) region (frequencies between 0.1–10 THz or energies 1-10 meV)[20]. In the current implementations, TCD is used more than TORD for the measurement of chiral phonons. In comparison between these techniques and ROA, TCD can provide unique structural and dynamical insights for biomolecules, molecular crystals and nanoscale assemblies because the sign, intensity and position of TCD peaks are highly sensitive to the three-dimensional organization of matter. Moreover, the ability of TCD to resolve low-frequency vibronic transitions offers a new window into phenomena such as chiral phonons and their propagation in nanostructured matter.

Unlike ECD and VCD, which use continuous-wave light sources and optical elements such as diffraction gratings and interferometers, TCD experiments employ pulsed THz radiation. In THz time-domain spectroscopy (THz-TDS) a femtosecond laser excites a photoconductive antenna or nonlinear crystal (e.g., ZnTe, GaP, GaSe) to generate a short THz pulse; a time delayed femtosecond probe pulse samples the emitted THz field in the time domain, enabling coherent detection of the THz waveform. By scanning the delay between pump and probe pulses, the transient THz electric field is recorded in the time domain, providing direct access to the amplitude and phase of the THz waveform in the horizontal and vertical directions. The polarization dependence of the emitted field, measured with and without the sample, can thus be extracted and converted to frequency-domain spectra through Fourier transformation. This direct measurement of the electric-field vector differentiates THz-TDS from Fourier-transform infrared spectroscopy (FT-IR), where only intensity is measured.

Substantial difference is also observed in respect to polarization modulation and optics: In ECD and VCD measurements, circular polarization is generated with photoelectric modulators or quartz quarter-wave plates, but such modulators are uncommon at THz frequencies because standard birefringent materials (fused silica, calcite, $MgF_2$, ZnSe) do not operate effectively in the THz range. Consequently, TCD requires specialized optical setups. Early implementations for metamaterials employed a three-polarizer arrangement: two fixed wire-grid polarizers (P1 and P3, **Fig. 3a**), which define the polarization states for the THz generation and detection, and a third polarizer (P2) positioned between the sample and detector that is rotated to collect the orthogonal field components ($E_x$ and $E_y$). The difference between the signals recorded at WGP2 are orthogonal to each other, removing instrument-dependent anisotropies and yielding the polarization rotation and ellipticity imparted by the sample. However, this arrangement requires rotation of the sample or polarizer between measurements; rotation is acceptable for metamaterials with periodic structures but can introduce large artefacts when applied to biomaterials or nanostructures where the beam spot (hundreds of micrometers) covers only a few unit cells. To address the rotation issue, researchers have developed several options for dynamic polarization modulators. One example is a chiral plasmonic kirigami metasurface inspired by paper-cut art, which consists of slanted Au strips on a flexible substrate[82]. Mechanical stretching modulates the local geometry of the kirigami sheet and tunes the polarization rotation and ellipticity of the transmitted THz beam up to~80° and 40°, respectively. This metasurface acts as a broadband, enantiomerically switchable quarter waveplate and allows static samples to be probed without physical rotation. The authors note that such modulators are crucial for TCD spectroscopy of biomolecular and nanoscale samples.

Among challenges we need to highlight ongoing optimization of sample requirements and sensitivity that is specifically important for chiral phonons. The long wavelengths of THz radiation (~300 µm) mean that focused beam spots are large and that photon–phonon interactions are weak. Consequently, large volumes of sample are often needed to obtain a measurable TCD signal. Water absorption in the THz range is also a major challenge; biological specimens must be dried or measured in deuterated solvents to reduce attenuation and TCD peak broadening. Because the THz signal power levels are typically low[83], TCD data often exhibit a high noise level. Values of polarization rotation or ellipticity below 0.5 degrees are regarded as noise, and thus sensitivity enhancements, such as using plasmonic nanostructures that concentrate the THz field or cryogenic detectors, are active research areas. We suggest that combining TCD with THz imaging modalities can further suppress artefacts and improve signal-to-noise by analyzing spatial variations across a sample using a hyperspectral, and potentially a machine learning-based approach[84] (**Fig. 3b**).

Because TCD measures the differential absorption of RCP and LCP THz radiation, it is well suited to resolve left- and right-handed chiral phonons. In fact, TCD can distinguish multiple overlapping phonon bands by probing differences in the complex refractive index (real and imaginary parts) of a sample, which ordinary THz absorption lumps into a broad featureless band (**Fig. 3c**). For example, simulations of an α-helical polypeptide predicted that TCD and TORD spectra would exhibit distinct positive and negative peaks at 2.24, 3.44 and 4.04 THz, whereas the corresponding THz absorption spectrum decays monotonically with no sharp features[20,85].The ability of TCD to resolve these peaks underscores its potential to study conformations and dynamics of macromolecules that lack clear absorption signatures. Experimental TCD spectra confirm the sensitivity of the technique. Using polarization modulators, researchers measured the TCD spectra of several biological samples, including the elytron of a green beetle, leaves of sugar maple trees, dandelion petals, and crystalline peptides, without rotating the sample. L-cystine crystals displayed mirror-image TCD spectra for L- and D-enantiomers at ~0.73 THz, demonstrating that TCD can resolve chiral phonon modes associated with specific amino acids. Distinctive positive-negative peaks were also observed in microcrystals of histidine, glutamine, glutamic acid, threonine and tyrosine, with the peak positions correlating with the symmetry of their crystal space groups[20]. Studies on suspensions of crystalline amino acids showed that chiral phonons persist even in slurries, and the associated sharp TCD and TORD peaks were recorded alongside THz absorption spectra. These observations indicate that chiral phonon signatures are intrinsic to the molecular organization rather than artefacts of crystal packing. By measuring TCD or THz Raman optical activity in these materials, one can probe chiral phonons that are both IR- and Raman-active, while taking advantage of the unique spectral measurement capabilities afforded by the two techniques. Such studies using TCD and circularly polarized Raman spectroscopy have recently been performed on enantiomers of amino acids, revealing commonalities as well as differences in the low-frequency chiral phonon modes in crystallized samples[86].

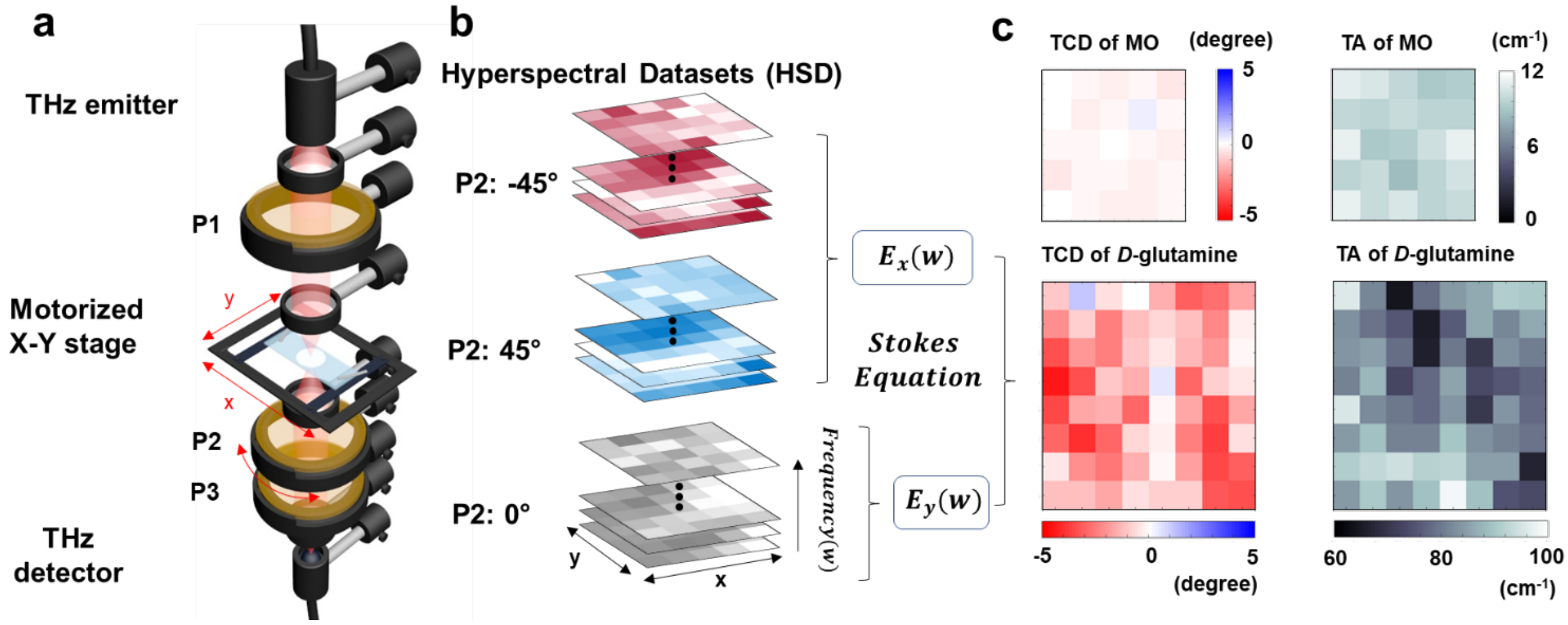

**Figure 3. Practical implementation of TCD and TORD chiroptical spectroscopies**. (a) The schematics of the experimental set-up. (b) Schematics of extraction of chiroptical spectra using the Stokes equation. (c) Example of the experimental data for mineral oil (MO) and *D*-glutamine.

By measuring TCD or THz Raman optical activity in these materials, one could probe valley phonon Hall effects and topologically protected edge modes in 2D nanomaterials[87]. 2D chiral phonons are interesting for photonics applications because they are strongly sensitive to chemical composition, interlayer interactions and structural symmetry. An important example of chiral phonons in 2D materials is the Archimedean spiral meta-surface[88], a kirigami sheet with slanted metallic strips can modulate the polarization rotation and ellipticity of a THz beam, and 3D printed helical arrays[89], enabling the detection of chiral phonon signatures at frequencies up to 80° polarization rotation[90]. TCD spectra recorded with such metasurfaces revealed strong phonon–photon coupling and allowed identification of valley chiral phonons. These metamaterial platforms also highlight the possibility of strong coupling between THz photons and chiral phonons, forming polaritons. In polaritonic systems, coherent Rabi oscillations and splitting in the THz range have been observed for lactose crystals[91]. The authors suggest that judiciously tuning the THz field to match the phonon frequency enables polaritonic chemistry, manipulating molecular reactions by controlling hybrid phonon–photon modes.

*Transient techniques*

The functionality of chiral phonons is often illustrated by their non-equilibrium dynamics, i.e., when the phonons evolve in space and time and interact with other degrees of freedom in materials. Ultrafast transient spectroscopy is a powerful approach to create and track non-equilibrium chiral phonons on their inherent timescales, femtoseconds to nanoseconds. Here, we systematically categorize excitation and probe methods in transient chiral phonon spectroscopy into *asymmetry-enabled* interactions, in which the phonon chirality arises from the inherent broken symmetry in a chiral material, and *symmetry-breaking* interactions, in which the phonon chirality is selected in symmetric materials by external polarization control (**Fig. 4a**). Both methods can be further divided into two categories: coherent (generating or measuring the displacement field) and incoherent (generating or measuring the phonon population). Once present in the materials, non-equilibrium chiral phonons can exhibit non-trivial transport properties or convert to electronic excitations, spin, and heat, whose dynamics are observable by a variety of time-resolved techniques.

Symmetry-breaking processes are obviously interesting because they inject additional information into the system, enabling a given material to behave differently as needed for device applications. At the same time, asymmetry-enabled processes are not trivial either, as they permit the transduction and amplification of symmetry properties across different entities. On the one hand, in a chiral material without any structural mirror symmetry, non-equilibrium states generally do not possess mirror symmetry. Therefore, a transient population of optically induced phonons in such materials, regardless of their conversion mechanisms, is expected to be structurally chiral. Because a directional light injection breaks time-reversal symmetry, these phonons may also break time-reversal symmetry and carry AM. On the other hand, a magnetic material that exhibits finite AM, together with an incident light beam carrying linear momentum in the same direction, is dynamically chiral[11]. Thus, phonons generated by light-matter interaction in such a system may be chiral as well. In many experimental cases, physical observations directly relate to the AM of chiral phonons (i.e., axial phonons), whereas their structural chirality is purely conceptual and has limited physical consequences. Nevertheless, one may also envision the opposite scenario in which the chemical effects of these light-induced phonons are prominent, but the time-reversal symmetry-breaking dissipates quickly and does not affect the end products.

Incoherent asymmetry-enabled process

Incoherent chiral phonons are usually excited in materials thermally and probed by phase-insensitive scattering, which may exhibit polarization selection proportional to their AM (**Fig. 4b**). An early conjecture of symmetry transduction through phonons was given by time-resolved elliptical photoluminescence (PL) in ferromagnet-semiconductor hybrid structures[92]. A long-range proximity effect appears between a magnetic layer of cobalt and a semiconducting quantum well of CdTe separated by a large-gap (Cd,Mg)Te spacer. The spin polarization of acceptor-bound holes slowly decays but remains robust over spacer thicknesses exceeding 30 nm. Here the AM was thought to arise from elliptically polarized phonons that naturally exist in thermal equilibrium in the Co layer and penetrate the spacer to reach the quantum well. After linearly polarized optical excitation, which does not alter the system's symmetry, light emission revealed the spin polarization of injected holes after 2 ns under phonon-mediated exchange coupling. Here, because the system already breaks time-reversal symmetry, transient spectroscopy is important for distinguishing various mechanisms, such as magnetic stray field and spin-dependent electronic capture, from phononic a.c. Stark effect.

The transfer of broken symmetry can also manifest in time-resolved magneto-optic Kerr (TR-MOKE) measurements. An anomalously long-lived Kerr rotation was observed in oxide superlattices of $SrRuO_3$ and $SrTiO_3$ near the Curie temperature of the ferromagnetic transition[93]. An additional sub-nanosecond remagnetization signal emerges in superlattices with various thicknesses compared with the single-component thin films, indicating strong coupling of the $SrRuO_3$ with or through the lattice of $SrTiO_3$. The dynamics clearly separated the thermal demagnetization, order recovery, and the relatively slow propagation of the spin-phonon coupling (an order of magnitude smaller than sound velocity), offering many important details for future theoretical studies. Spin-induced chiral phonons are expected to propagate at different rates in chiral materials, as observed by time-domain thermoreflectance[94]. Inversely, the CPASS effect allows chiral materials to induce spin dynamics in metals measurable through TR-MOKE[19].

Another indirect piece of evidence for spin-to-phonon symmetry transfer was provided by ultrafast electron diffraction (UED) in single-crystal nickel[95]. Following laser-induced demagnetization, a non-equilibrium population of anisotropic high-frequency phonons quickly develops within 150 fs - 750 fs. These phonons oscillate in a plane perpendicular to the magnetization vector incoherently, reducing the intensity of the diffraction spots in the plane. Unfortunately, UED cannot directly measure the AM of incoherent phonons. Therefore, achiral processes that could produce the same observables, such as magnetostriction, were ruled out by an order-of-magnitude estimate.

Coherent asymmetry-enabled process

Femtosecond ultrafast probes are sufficient to resolve the displacement fields of phonons, which modulate the inherent chirality of materials, and directly prove phonon chirality. For example, in thin ferromagnetic iron films, ultrafast demagnetization triggered by a laser pulse transfers AM to acoustic phonons, generating a transient torque on surfaces parallel to the initial spin orientation[96]. This torque launches a coherent transverse strain wave from the interface into the bulk, accounting for approximately 80% of the lost spin AM within 200 fs. Consequently, time-resolved X-ray diffraction observed an oscillatory diffraction intensity expected to change sign for opposite initial magnetization.

Coherent acoustic phonons excited by ultrafast pulses in chiral materials not only cause periodic transient reflectance and absorption, like in a typical photoacoustic process, but also oscillations in transient circular dichroism (TrCD) and impulsive stimulated Raman spectroscopy. Following an optical pulsed excitation without external symmetry breaking, CD oscillations were observed in chiral polyfluorene copolymers, and their frequency is determined by the longitudinal sound velocity[97]. Similarly, coherent optical phonons excited by the electric field of a THz pulse modulate the "chirality amplitude" in a chiral crystal $CsCuCl_3$ (**Fig. 4c**)[98]. Although there is not yet a satisfactory way to quantify structural chirality from the phonon eigenmode and displacement amplitude, in resonant X-ray diffraction of $CsCuCl_3$ the (1020) peak becomes visible and circular dichroic only below the chiral phase transition. Thus, an oscillation in the circular contrast of this peak at a phonon frequency indicates that the atomic displacements of this phonon mode are chiral.

Incoherent symmetry-breaking process

In materials without AM or structural chirality, selective interaction with chiral phonons and/or induction of chirality may be achieved by an incoherent circularly polarized pump. The earliest experimental demonstration of chiral phonons was performed in a non-magnetic, achiral material, monolayer semiconductor $WSe_2$[7]. The optically injected holes are spin-orbit polarized, which then enable valley-selective indirect absorption involving valley chiral phonons (the longitudinal optical phonons at the K-point) and showing transient infrared circular dichroism (**Fig. 4d**). The temporal dynamics occurred because of spin-orbit polarization rather than the LO(K) phonons, and the results proving the coupling between valley-polarized excitons and phonons were later confirmed by static PL measurements[99–103]. Valley phonon dynamics after optical excitations were observed by momentum-resolved ultrafast electron diffuse scattering, as well as transient optical absorption under intense optical excitation in the nonlinear phononics regime, but unfortunately, the excitations were not circularly polarized for resolving phonon chirality[104,105].

Circularly polarized excitation can also distinguish phonons in a mixture of unresolvable chiral domains in apparently achiral materials with possible chiral order. For example, polarization-contrast lattice dynamics was demonstrated for chiral charge density waves (CDWs) in 1T-$TiSe_2$ using time-resolved X-ray diffraction (tr-XRD).[106] Left- and right-circularly polarized excitation at 800 nm results in a 20% difference in the suppression of the CDW diffraction peak intensity, attributed to preferential excitation of carriers within chiral domains. The incoherent structural dynamics imply an instantaneous increase in the CDW correlation length and may offer a route to "chiral training", i.e., the preference of one structural chiral domain under polarized illumination through chiral electron-phonon coupling.

Coherent symmetry-breaking process

Advances in laser spectroscopy allow phase-locked coherent chiral phonons driven by circularly-polarized stimulated Raman scattering and resonant THz absorption. Early demonstrations of rotational atomic trajectories were done in α-quartz, which are structurally chiral, but the approach is applicable to any crystal with 2- or more-fold rotational symmetry[107]. In contrast to single-pulse-generated coherent phonons discussed in asymmetry-enabled processes, a pair of linearly polarized pulses, differing by 45° and with a varying time delay, controls the polarization of degenerate, Raman-active E-symmetry optic phonons. A delay corresponding to $\pm\pi/2$ phase of the phonon produces clockwise and counterclockwise motion, causing rotational birefringence that is fully determined by the ellipticity measurements of two optical probes. Alternatively, Raman-active chiral phonons are addressable by infrared pulses via nonlinear phononics and AM conservation[108].

More recently, the development of broadly tunable circularly polarized THz sources has enabled resonant excitation of infrared-active chiral phonons and the study of their dynamics in details[18,109]. The chirality of phonons is observable using conventional techniques, such as time-resolved Raman scattering or electro-optic sampling, in materials with broken inversion symmetry. But for symmetric materials, nonlinear spectroscopy, such as time-resolved hyper-Raman scattering and electric-field-induced second-harmonic generation, is required[110]. In either case, time-domain methods allow direct extraction of the motion trajectory, while frequency-domain methods may lose phase information unless a heterodyne detection scheme is implemented with a reference local oscillator. Comparing coherent and incoherent dynamics in achiral materials reveals the loss mechanisms of phonon chirality. For example, in monolayer $WS_2$, the two dynamics obey a simple relationship $T_2 = 2T_1$, indicating fast conversion between left- and right-handed phonons by lattice strain[109].

Coherent chiral phonon fields driven by intense THz pulses are crucial in proving strong phonon-induced magnetic effects, in which the effective magnetic field is determined by the net AM and is not phase-sensitive. In magnetically ordered materials, magnetoelastic coupling allows atomic displacements to modulate spin exchange, leading to linear magnon-phonon hybridization and nonlinear ionic Raman scattering, but the spin dynamics are usually dictated by the material's symmetry rather than the polarization of phonons[62,111]. The first observation of the chirality-driven magnetic effect was reported in noncolinear antiferromagnetic orthoferrite $ErFeO_3$, in which the simultaneous excitation of orthogonal infrared-active phonons with controlled relative phases creates an elliptical lattice polarization[112]. The phase of the resulting coherent spin precession is governed by the phonon ellipticity, corresponding to an effective magnetic field proportional to the ionic AM, reaching 36 mT for a fluence of 20 mJ/cm$^2$.

Surprisingly, a much stronger magnetic effect from chiral phonons was recently observed by magneto-optic spectroscopy in paramagnetic rare-earth halide $CeF_3$, boosting the field strength to 0.9 T under a mere fluence of 0.44 mJ/cm$^2$ (Ref. [18]). The field strength was quantified by carefully modeling the paramagnetic relaxation dynamics and calibrating the magneto-optic coefficient, and was shown to agree well with the expected phonon inverse Faraday effect (**Fig. 4e**)[32]. The giant spin-phonon coupling in rare-earth halides originates from the near-resonant coupling between phonons and crystal electric field levels, but its magnitude is not fully explained[113]. More unexpectedly, evidence of phonon magnetic effect emerges in non-magnetic materials, such as the quantum paraelectric perovskite $SrTiO_3$ and ferroelectric $LiNbO_3$[114,115]. Even in simple insulators like sapphire ($Al_2O_3$) and glass ($SiO_2$), the effective field appears to penetrate a 5-nm $Si_3N_4$ spacer and is large enough to reverse the magnetization of the GdFeCo film on top[116].

Finally, it is worth noting that “geometric chiral” and “axial but not chiral” atomic movements have been experimentally demonstrated by coherent THz excitation. Deterministic induction of structural chirality was realized in $BPO_4$ nonlinear phononic rectification, in which an intense THz pulse drives an infrared-active E symmetry mode, exerting a displacive force on an inactive B mode[117]. Since the frequency of the B mode is much higher than the impulsive rectification, no coherent phonon oscillation is excited, but a unidirectional chiral displacement is found by optical activity. Likewise, deterministic switching of ferroaxial order is reported in $RbFe(MoO_4)_2$, where the nonlinear interaction between the THz electric field and the resonantly excited infrared-active mode presses the optically inactive axial mode[118]. The axial displacement is unidirectional and transient above the ferroaxial Curie temperature, but frozen below, enabling single-pulse switching verified by circular dichroic second harmonic generation.

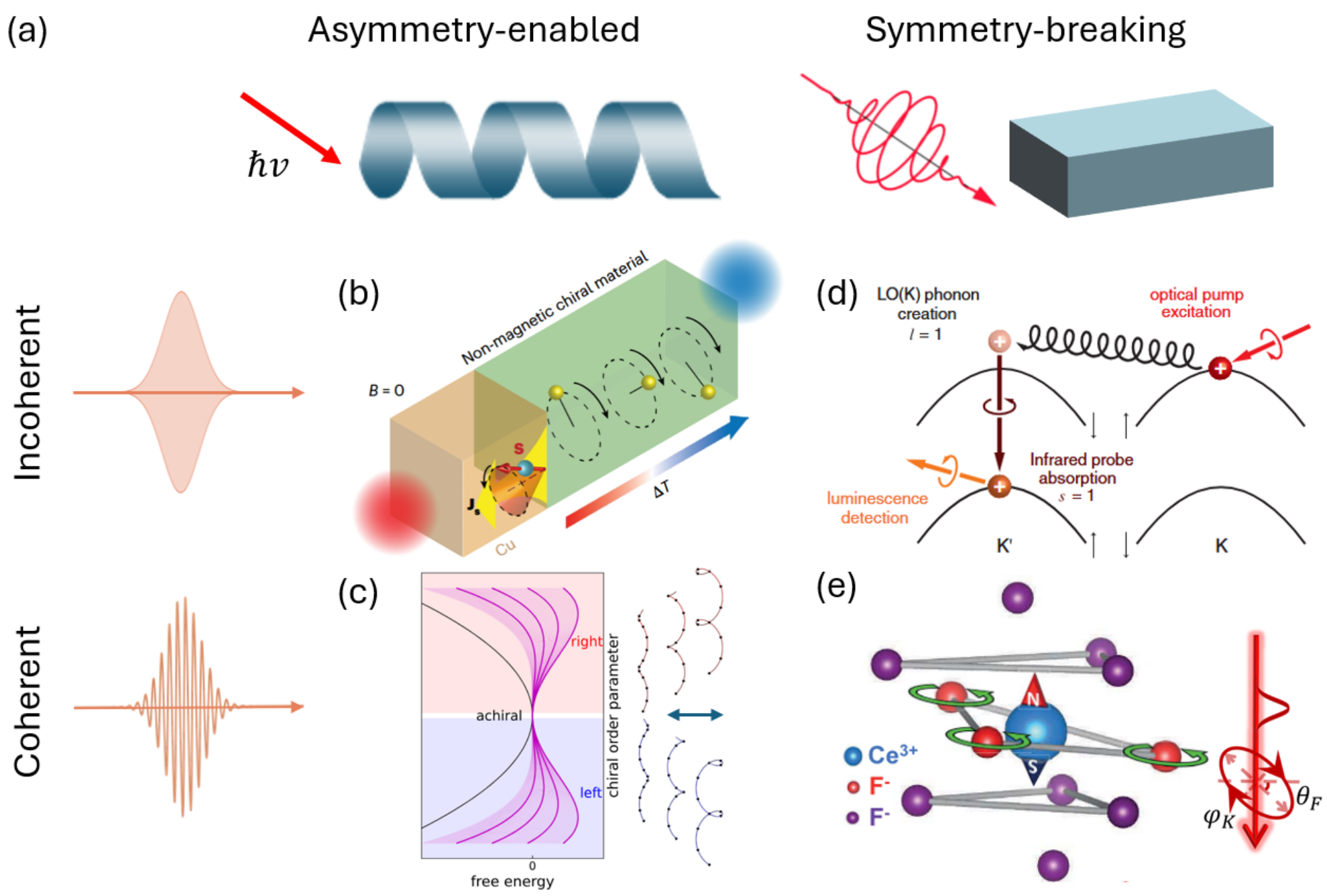

**Figure 4**. **Generation and detection of non-equilibrium chiral phonons in materials through ultrafast light-matter interaction.** (a) When chiral and magnetic materials interact with incoming light, the system configuration is chiral. Light-induced chiral phonons are distinguished from other chiral processes by temporal dynamics and signal magnitude. In contrast, polarized light can select the handedness of the transient state in achiral materials. (b) The spin transfer torque in chiral-phonon-activated spin Seebeck (CPASS) effect is caused by incoherent thermal phonons and determined by the chirality of the organic-inorganic hybrid perovskite. From Ref. [19] (c) Coherently oscillating ellipticity in resonant X-ray scattering measures the phonon displacement along the achiral-to-chiral phase transition coordinate. From Ref. [98] (d) Infrared circular dichroism proves the chirality of phonon-assisted indirect absorption in monolayer $WSe_2$, whose valley symmetry is broken by circularly polarized optical excitation. From Ref. [7]. (e) Coherent chiral phonons resonantly excited by circularly polarized THz pulses spin-polarize the paramagnetic $CeF_3$. From Ref. [18].

**BOX 2**

**Application venues for chiral phonons**

Biosensing

The strong and narrow THz spectral bands from chiral phonons offer an almost 3 orders of magnitude improvement over the ECD of molecules in the visible spectrum owing to their strong association with the crystal structure and intermolecular bond strength[20]. This sensitivity translates directly into quality control of drugs and supplements, such as detecting subtle manufacturing batch variations in the dipeptide *L*-carnosine. By combining strong amplitudes with partial transparency of biological tissue to THz radiation, chiral phonons have also been harnessed for biological imaging, such as identifying cystine-based bladder stones[20] and tracking drug dissolution through strong chiral phonon-plasmon resonance[21].

Spintronics

Recent experiments[19,119,120] have established that chiral phonons enable spin–thermal conversion, through the chiral phonon–activated spin Seebeck (CPASS) orbital Seebeck effect (CPOSE) effects. Here, a temperature gradient across an interface between a nonmagnetic material and a metal drives chiral phonons out of equilibrium, transferring AM across the interfaces to generate spin or orbital currents.[121,122] This is possible even in the absence of a magnetic order or external magnetic fields. Experiments on ferroelectric triglycine sulfate also demonstrated electric field control oh phonon chirality.[123] Hence, electrical and thermal control over chiral phonons allows spin/orbital current generation in functional nonmagnetic systems, opening promising avenues for spin caloritronics, chirality-based information encoding, and nonreciprocal spin–heat devices.

Chiral phonon-selective chemistry

In chiral solids, preferential populations of phonons with one handedness or the other can guide reaction products, advancing enantioselective chemistry[124]. A specific promising research direction is the excitation of a catalyst surface with (ultrafast) circularly polarized light, resulting in chiral phonons that can aid in enantioselective chemical bond formation through the preference of handed atomic geometry or spin alignment taking part in the reactions. Theoretically, since the handedness of chiral phonons can be reversed in real-time by reversing the circular polarization of the excitation[125], this potentially offers a degree of control over chiral products. Indeed, recent theoretical studies[121] have indicated that chiral phonons may play a significant role in the chirality-induced spin selectivity (CISS) effect[126]. Thus, the flow of excitations through chiral solids[9,127,128] can offer new pathways to control spin in chemical processes, particularly those involving triplet states.

Thermal transport

Chiral phonons introduce new opportunities for controlling thermal transport through the AM of lattice vibrations. Unlike conventional phonons as heat carriers, their handedness allows directional and nonreciprocal heat flow, functioning as thermal diodes[129]. Additionally, chirality-selective magnon-phonon coupling in magnetic materials yields tunable anomalous Hall responses[130]. Tuning chiral phonon populations via composition, strain, or optical excitation can thus open new pathways for directional heat steering[129], phononic circuit elements[131], waste heat harvesting[19], and ultrafast thermal management[132] without charge transport.

Quantum information science

The discrete, well-defined AM carried by chiral phonons makes them natural candidates as information carriers in quantum systems because phonons can propagate over mesoscopic distances while remaining relatively isolated from electromagnetic noise. In 2D materials like transition metal dichalcogenides, opposite chirality at the *K* and *K*' valleys can enable qubit encoding[129], extending valleytronics into phononics. Quantum information could also be initialized optically, manipulated through electron-phonon coupling, and read out through their distinct spectroscopic signatures in diffuse scattering, optical absorption, or emission[100,133]. This ability to transduce quantum information between optical photons, excitons, and chiral phonons in a single material platform is particularly attractive for hybrid quantum architectures.

## Conclusions and Outlook

Chiral phonons have recently gained significance across diverse fields and applications including spintronics, spin-selective chemistry, thermal transport, and quantum information processing, yet the experimental communities probing them have largely worked in isolation. In this perspective, we surveyed the principal techniques for their detection, namely, circularly polarized Raman spectroscopy, THz circular dichroism, resonant inelastic X-ray scattering, and transient spectroscopy. We described their physical bases, implementations, and the distinct aspects of phonon chirality each can access, with the aim of providing a cross-disciplinary experimental roadmap.

Each technique has its strengths and limitations. Circularly polarized Raman and ROA access zone-center phonon modes with chemical specificity, but the chiral differential cross-section is approximately three orders of magnitude smaller than normal Raman intensities. The measurement requires highly uniform laser polarization states and avoidance of resonance conditions, which are sensitive to electronic rather than vibrational chirality. The frequency splitting between chiral phonon branches in optical scattering is limited by the momentum transfer, which is difficult to resolve in common commercial spectrometers. RIXS offers larger momentum transfer, but its resolution is also lower and comes at the cost of access to synchrotron sources. TCD uniquely resolves low-frequency collective modes and distinguishes overlapping phonon bands through their differential complex refractive index. Its principal constraints are practical: the long wavelength of THz radiation produces large focal spots and demands bulk sample volumes, standard birefringent optics fail in the THz range requiring custom polarization architectures, water absorption severely attenuates signals from hydrated samples, and the noise floor of ~0.5° polarization rotation masks weak chiral signatures in many systems of interest.

Transient spectroscopic methods such as coherent THz excitation, TR-MOKE, and ultrafast electron diffraction, can resolve non-equilibrium chiral phonon dynamics on femtosecond timescales, but disentangling phonon chirality from competing processes such as magnetostriction, orbital polarization, optical non-linearity and lattice strain remains a central interpretational challenge. Since tabletop optical methods mainly probe phonon chirality near the Brillouin zone center, it is crucial to develop ultrafast X-ray and electron-based techniques for accessing phonon chirality across the entire Brillouin zone. For example, femtosecond X-ray transient grating, currently being developed in free electron laser facilities, will directly inject coherent, large-momentum chiral phonons enabled by crystalline asymmetry. Combining multi-probe characterizations, such as time-resolved X-ray diffraction and polarized photoemission, one can directly observe and separate the structural and electronic dynamics following non-equilibrium chiral-phonon excitation. ```

The primary unmet need of the field is the convergence of the many techniques discussed herein. Interrogating the same material with circularly polarized Raman, THz-CD, and transient spectroscopy in tandem would test whether the phonon populations accessed by each technique are mutually consistent, resolve long-standing ambiguities in mode assignment, and provide the experimental benchmarks against which theoretical predictions of PAM can be validated. Several developments may accelerate this. For example, chiral plasmonic metasurfaces can eliminate the sample-rotation requirement that introduces artifacts in THz-CD; polariton platforms can enable coherent control of phonon handedness through photon–phonon hybridization; and the recent availability of a chiral phonon material database[33] provides a rational basis for selecting target systems across the 65 chiral space groups and beyond. In the meantime, a unified theoretical framework to quantitatively predict the chiral activity of phonons and/or their interactions with light and spin remains out of reach. What the field now requires is the collective instrumentation to measure a variety of materials with the precision and reproducibility demanded by functional exploitation for data-driven understanding and optimization of phonon chirality.

## Acknowledgements

R.R. and T.T.M. are grateful for support from the U.S. Air Force Office of Scientific Research (AFOSR) and from the Materials and Manufacturing Directorate, Air Force Research Laboratory.

H.Z. acknowledges the support from the National Science Foundation (DMR-2240106) and the Welch Foundation (C-2311).

The work of W.C. was performed under the auspices of the U.S. Department of Energy by Lawrence Livermore National Laboratory under Contract No. DE-AC52-07NA27344.

R.R.F. acknowledges support by the AFOSR under Grant No. FA9550-23-1-0368.

D.S. and J.L. acknowledged the support by the AFOSR, Multidisciplinary University Research Initiatives (MURI) Program under award number FA9550-23-1-0311.

## Author contributions

The authors contributed equally to all aspects of the article.

## Competing interests

The authors declare no competing interests.